\documentclass[a4paper,12pt,reqno,superscriptaddress]{revtex4}
\usepackage{graphicx}
\usepackage[centertags]{amsmath}
\usepackage{amsfonts}
\usepackage{amssymb}
\usepackage{amsthm}
\usepackage{newlfont}
\usepackage{stmaryrd}
\usepackage{mathrsfs}
\usepackage{euscript}

\usepackage{color}

\theoremstyle{plain}

\theoremstyle{definition}

\theoremstyle{remark}

\numberwithin{equation}{section}

\newcommand{\bbR}{{\mathbb R}}

\newcommand{\opunit}{\text{1}\kern-0.22em\text{l}}

\DeclareMathAlphabet{\mathpzc}{OT1}{pzc}{m}{it}

\newcommand{\id}{\textrm{d}}

\begin{document}

\title{On the second fluctuation--dissipation theorem\\ for nonequilibrium baths}

\author{Christian Maes\\
Instituut voor Theoretische Fysica, KU Leuven,
Belgium} 

\begin{abstract}
Baths produce friction and random forcing on particles suspended in them.
The relation between noise and friction in (generalized) Langevin equations is usually referred to as the second fluctuation-dissipation theorem.   We show what is the proper nonequilibrium extension, to be applied when the environment is itself active and driven. In particular 
we determine the effective Langevin dynamics of a probe from integrating out a steady nonequilibrium environment.  The friction kernel picks up a frenetic contribution, i.e., involving the environment's dynamical activity,  responsible for the breaking of the standard Einstein relation.

\end{abstract}

\maketitle




\section{Reduced dynamics}
\begin{figure}[t]
\hspace{-1cm}\includegraphics[angle=-90,width=11 cm]{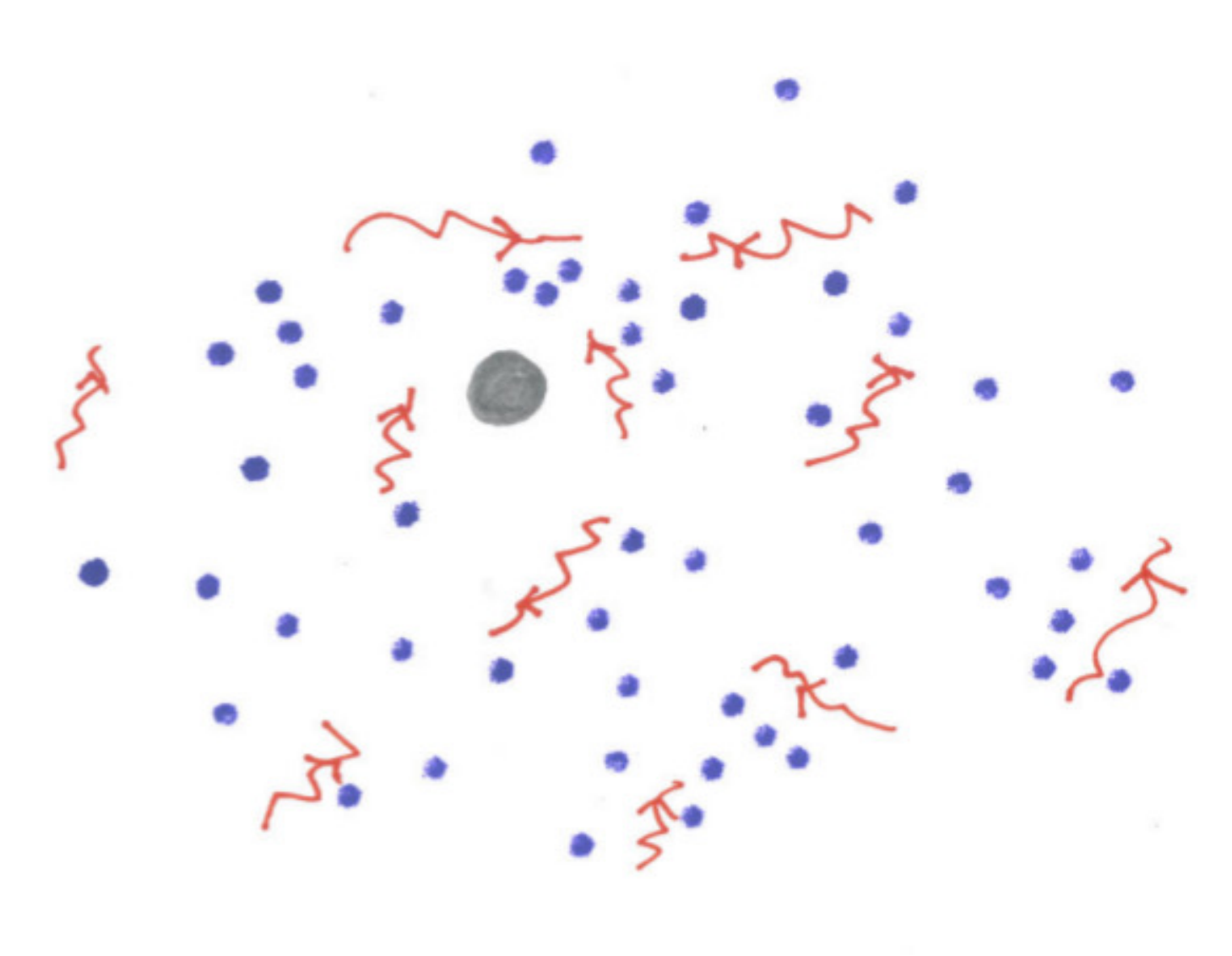}
\caption{Three levels, a probe interacting with driven particles in a heat reservoir.  The probe can for example be a silica bead attached to the cytoskeleton of a living cell pushed and carried around by molecular motors and with a thermal bath as background medium. The question is to describe the effective diffusive behavior and mobility of the probe.}\label{on3}
\end{figure}

The Langevin equation or its generalizations are effective diffusive dynamics to describe certain tagged degrees of freedom interacting with a heat bath.  Best known is the case of a Brownian particle suspended in a fluid at rest.  In a growing number of applications the tagged particle or probe moves in a driven or active medium of particles, the latter in turn being in contact with an equilibrium heat bath. In that way there are three levels of description: probe, driven particles and heat bath --- see Fig.~\ref{on3}. For the driven particles we have in mind active media such as the cell environment of a living organism in which the motion of microprobes is studied \cite{pro,ros}, or spatially extended objects such as large polymers undergoing nonequilibrium forcing and for which the motion of a tagged monomer is investigated \cite{rouse,gup}. We can also imagine a sheared or non-uniformly rotating and thermostated fluid in which colloids or polymers are moving; see e.g. \cite{santa,depl,holz,drien} among many possible references. In fact, baths can be out-of-equilibrium for a great variety of reasons.  Here we do not concentrate on one special case but go for the general structure of the effective dynamics of a probe in weak contact  with many constituents under steady driving.  For better focus we do not consider the effect of time-dependent reservoirs like periodically driven charged matter, and we are also not concerned with quantum aspects; see e.g. \cite{rig, cliv} for such situations.\\

The reduced or effective time-evolution of the probe is obtained by integrating out all other degrees of freedom.
 As under equilibrium conditions, one expects a definite relation between the friction and the noise in the reduced evolution equation.  After all, noise and friction terms have the same physical origin in the collisions or more generally in the interaction with the hidden particles.  Traditionally, that relation follows the second fluctuation-dissipation theorem, giving a proportionality between the noise amplitude and the friction kernel.  We review the origin of that relation in Section \ref{2nd}, but invariably, some assumption of (local) equilibrium of the bath is involved.  Hence the question of the present paper, what is the proper extension and modification of the second fluctuation-dissipation theorem when dealing with nonequilibrium baths?\\

It is important to keep the distinction with the {\it first} fluctuation--dissipation relation.  An example of the latter is the Green-Kubo formula which for a given thermodynamic context expresses the (linear) response coefficients as a time-integral of the current autocorrelation under the equilibrium dynamics. The Kubo formula in general gives the perturbed expectation in terms of the correlation function between the observable in question and the entropy flux due to the perturbation.
There are also modified or extended such fluctuation-dissipation theorems for driven particles, e.g. in contact with different equilibrium reservoirs. A short review of the most recent wave of results is available in \cite{up}. 
The goal of the present paper is to derive the nonequilibrium version of the second fluctuation-dissipation theorem from these (previously derived) nonequilibrium extensions of the (first) fluctuation--dissipation theorem, staying in line with the programme of unifying the influence of time-symmetric kinetic factors via the notion of dynamical activity. \cite{fren,fdr}.  That is being summarized in the frenetic contribution; the correlation between e.g. position and active forces plays a role there and becomes visible in the modified relations \cite{proc,pro,sog}.\\

The plan of the paper is  the following.  The next section starts with the main finding and discussion of the result.  The derivation is in Sextion \ref{n2}.  In between we present a reminder on three ways to derive the (traditional) second fluctuation--dissipation theorem.  We highlight there the  equilibrium input. 
Similarly, in Section \ref{nfdr} we state briefly the necessary ingredients from the linear response theory around nonequilibria.  We are then ready for the calculation in Section \ref{n2} to give the derivation of the nonequilibrium version of the second fluctuation--dissipation theorem, our main result.\\

Application to specific cases and models are postponed to another paper.  We add however here already that the question of the appropriate nonequilibrium generalization of the second fluctuation--dissipation relation goes quite beyond (generalized) Langevin dynamics. The real problem it addresses is the characterization of effective (or, statistical) forces well outside equilibrium and how they relate with the system's fluctuation behavior.

\section{Discussion of result}\label{disc}
The (standard) second fluctuation--dissipation theorem is not violated in an arbitrary way, but there exists a systematic modification related to further kinetic aspects of the active medium that already enter in the study of nonequilibrium linear response.\\

We start from a specific example to explain our findings.  
We take an oscillator model for the probe $q_t\in S^1$ and the medium (called sea) degrees of freedom  $Q_t=\{q^j_t\in S^1\}$ on the unit circle.  The sea spins $q^j_t$ are mutually independent but weakly coupled to the probe. Half of them are driven clockwise and  half of them are driven counter clockwise by the influence of a constant external field $F^j =  (-1)^j\,F,\, j=1,\ldots,N$:
\begin{eqnarray}\label{hamin}
\Gamma\,\frac{\id q_t}{\id t} + V'(q_t)&=& 
 - \lambda\,\sum_j \sin\,[q^j_t-q_t], \quad q_0=0\nonumber\\
\dot{q}^j_t &=& F^j + a\,\sin q^j_t
 + \lambda\,\sin [q^j_t-q_t] + \sqrt{2T}\,\xi^j_t
\end{eqnarray}
   We assume that $q_0=0$ is the preferred probe direction, with $V'(0)=0$ and $\Gamma$ is large to damp the oscillations around $0$.  
In fact the effective description that follows works well for $q_t=O(\lambda)$,  where  the coupling $\lambda$ is weak, and it is useful to consider the case $\lambda \propto 1/\sqrt{N}, \,N\uparrow \infty$.  The active spins $q^j_t$ undergo a conservative force of amplitude $a$ and are driven by $F$, either clockwise or counter clockwise, so that the nonequilibrium position of the probe $q_t$ is on average still decided by the periodic potential $V$.\\   

A first approximation to \eqref{hamin} is obtained by assuming that the relaxation of the sea is much faster than for the probe.  In that scenario of infinite time-scale separation, we can use the stationary density $\rho_{q_t}(q^j)$ of the sea spins for {\it fixed} $q_t$ in \eqref{hamin} to take the averaged probe dynamics
\[
 \Gamma\frac{\id q_t}{\id t} 
 + V'(q_t) = G(q_t)
 \]
 with statistical force
 \[
 G(q) =  - \lambda\,\frac{N}{2}\big[\langle \sin(\theta-q)\rangle^q_F + \langle \sin(\theta-q)\rangle^q_{-F} \big]
\]
where $\langle\cdot\rangle^q_{\pm F}$ is the stationary expectation for the dynamics
\[
\dot{\theta}_t = \pm F + a \sin\theta_t +\lambda\,\sin(\theta_t-q) + \sqrt{2T}\,\xi_t,\quad \theta_t\in S^1
\]
The force $G$ is conservative and vanishes at $q=0$.\\
A better approximation is to allow delay effects for the sea spins which lead to friction and noise.  
 We then find the effective dynamics of the probe to be
 \[
 \Gamma\frac{\id q_t}{\id t} + V'(q_t)= G(q_t) - \int_0^t \gamma_s\,\dot{q}_s\,\id s + \eta_t, \quad q_0=0
 \]
  For the friction kernel to order $\lambda^2$ we find
\begin{eqnarray}\label{kura}
\gamma(s) &&= \frac{\beta\,\lambda^2 N}{2} \big[\langle \sin \theta_0 \,;\, \sin \theta_s  \rangle^0_F
+ \nonumber\\
&& - \int_{-\infty}^0\id u \,\{F\,\langle  \cos \theta_u \,;\,\sin \theta_s \rangle^0_F + \frac a{2}\langle  \sin 2\theta_u \,;\,\sin \theta_s \rangle^0_F - T\, \langle  \sin \theta_u \,;\,\sin \theta_s \rangle^0_F\} \big]
\end{eqnarray}
where the connected time-correlations $\langle A;B\rangle^0_F = \langle AB\rangle^0_F - \langle A\rangle^0_F\langle B\rangle^0_F$ are in the stationary process for the decoupled driven dynamics
 \begin{equation}\label{fav1}
 \dot \theta_t = F + a\,\sin \theta_t
 + \sqrt{2T}\,\xi_t
 \end{equation}
 with $\xi_t$ standard white noise.\\
 The noise $\eta_t$ has mean zero and stationary covariance 
\[
\langle \eta_0\eta_s\rangle = \beta\,\lambda^2 N\, \langle \sin \theta_0 \, ;\,\sin \theta_s  \rangle^0_F
\]
In the case of detailed balance, $F=0$, we get that $\beta \,\gamma^{\text{eq}}(s) = \langle \eta_0\eta_s\rangle$ which is the standard second fluctuation--dissipation relation  between friction and noise.  For nonequilibrium there is a correction; see Fig.~2.   Even in the Markov limit and while the stationary probe dynamics $q_t$ appears time-reversible, the effective temperature for the fluctuations around $q=0$  is not $T$ when $F\neq 0$.\\ 
\begin{figure}[t]
\hspace{-1cm}\includegraphics[width=11 cm]{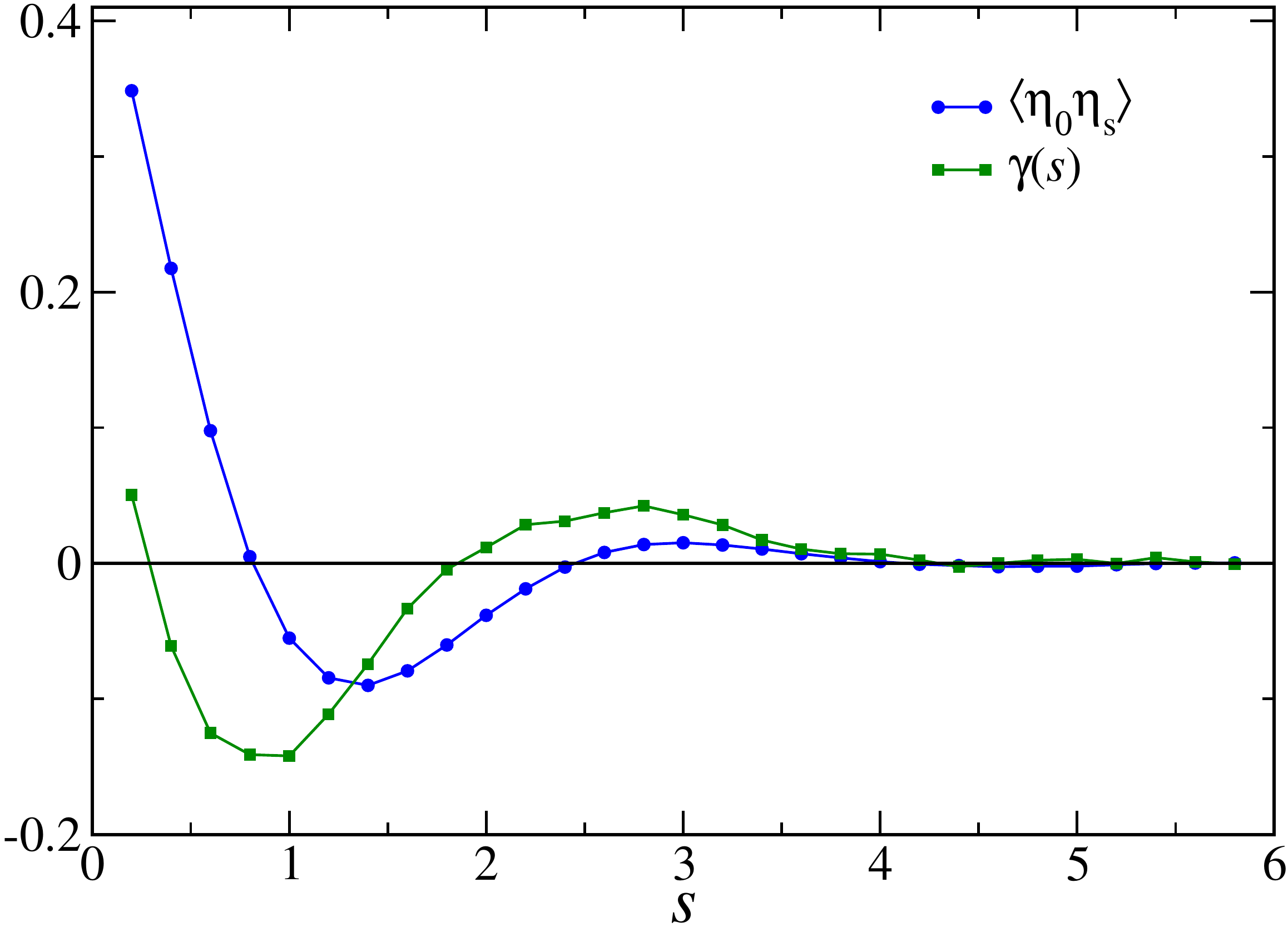}
\caption{The noise covariance and the friction memory kernel \eqref{kura} as function of time for the effective probe dynamics starting from the  nonequilibrium oscillator model \eqref{hamin} with $T=1$, $F=2$ and  $ a=1$.  Note that the values in the Markov limit $s\downarrow 0$ are positive, but not for finite memory.  When the oscillator sea is undriven, $F=0$, the two curves coincide exactly and show positive values for all times.  (Figure kindly provided by Urna Basu.)}\label{oscil}
\end{figure}

The result has the following structure. (Details follow in Section \ref{n2}.)   We consider an in  general high-dimensional variable $Q_t$ which denotes the state of the ``sea'' or active medium, and which we monitor via a probe variable $q_t$ of mass $M$ which corresponds to the slower degree of freedom.  For simplicity of presentation we restrict ourselves  to the simplest version of an overdamped dynamics in which $Q_t$ and $q_t$ are coupled via an energy function $U(Q_t,q_t)$ and we use ``one-dimensional notation.''  We then write for the evolution of the active degree of freedom
\begin{equation}\label{pertu}
\dot Q_t = -\partial_Q U(Q_t,q_t) + F(Q_t) + \sqrt{2T}\,\xi_t
\end{equation}
for non-conservative force $F$ and standard white noise $\xi_t$.  The prefactor to the noise contains the temperature $T$ ($k_B=1$) of the, for the rest invisible, heat bath in which $Q_t$ is immersed.  We want to integrate out the $Q-$degrees of freedom from the equation of motion for the probe
\begin{equation}\label{wh}
M\ddot q_t -  K_t(q_t,\dot{q}_t,\ddot{q}_t) = -\partial_q U(Q_t,q_t) 
\end{equation}
where the force $K_t$ is arbitrary and quite irrelevant for the discussion.  In the previous example \eqref{hamin} we took the overdamped limit.\\
The general set-up does not change whether $F=0$ of $F\neq 0$ in \eqref{pertu}.
Equilibrium or out-of-equilibrium, the integration over the bath degrees of freedom can be accompanied by a variety of limiting regimes, referred to in the literature as weak coupling, strong coupling, singular coupling, mean field coupling, adiabatic elimination, etc.  They are physically and conceptually sometimes very different, yet they do not necessarily lead to mathematically different reduced dynamics.  Of course what limit to take depends on the nature of the coupling between probe and sea, e.g. whether it is infrequent hard core collision or rather continued but weak and smooth interaction.  For the present paper we choose to consider a weak and smooth coupling where the resulting probe's position varies little around a fixed position $y$. (We took $y=0 $ in the previous oscillator example.)  We think of $y$ as installed and manipulated by the experiment, where the probe is trapped and we are asked to describe its fluctuating motion around it.  In particular, we do not consider the case where the probe would start moving in the sea because of a net current; see Section \ref{ar} for an additional remark.

Our result will be valid in linear order in the changes $q_s-y$, i.e., for small displacements of the probe and assuming that the coupling between probe and sea in the energy $U(Q,q)$ is sufficiently weak and smooth.  We then obtain (in Section \ref{n2}) the effective probe motion
\begin{equation}\label{who}
M\ddot q_t -  K_t(q_t,\dot{q}_t,\ddot{q}_t) = G(q_t) - \int_0^t\, \gamma(t-s) \,\dot{q}_s\,\id s + \eta_t 
\end{equation}
where we next specify the statistical force $G$, the friction kernel $\gamma(s)$ and the noise $\eta_t$.\\
The statistical force is the average force in the stationary density $\rho_q(Q)$ of the sea-dynamics \eqref{pertu} for fixed probe position $q_t=q$:
\begin{equation}\label{efff}
G(q) = - \int\id Q \,\rho_q(Q)   \,\partial_q U(Q,q)
\end{equation}
corresponding to the expectation of the right-hand side of \eqref{wh}.  In other words, the effective or statistical force $G$ is the force in the limit of infinite time-scale separation between sea and probe: at each probe position $q_t$ the sea relaxes instantaneously to the nonequilibrium density $\rho_{q_t}$.  That is in general the first term in the effective evolution after adiabatic elimination and it would be natural to suppose that $G(y) + K_t(y,0) =0$. In the case of detailed balance, $F=0$ in \eqref{pertu}, the statistical force $G(q)= -\partial_q{\cal F}$ is the gradient of the sea equilibrium free energy; we come back to that case under \eqref{supper}.\\
 The next order brings both friction and noise.  Write $g(Q,q) = - \partial_q U(Q,q)$ for the mechanical force of the sea on the probe which is assumed small, say of order $\varepsilon$.  Then, from \eqref{efff}, $G$ is of order $\varepsilon$ as well.  The friction kernel is of order $\varepsilon^2$:
\begin{equation}
\label{fric}
\gamma(s) = \frac{\beta}{2}\big[\langle g(Q_0,y)\,;\, g(Q_s,y) \rangle^y - \int_{-\infty}^0\id u \langle Lg(Q_u,y) \,g(Q_s,y) \rangle^y\big],\quad s\geq 0
\end{equation}
where the expectations $\langle \cdot \rangle^y$ are with respect to the sea dynamics \eqref{pertu} while fixing the probe $q_t = y$ at its preferred position, and $L$ is its backward generator
\[
Lg(Q,y) = (F(Q)-\partial_Q U(Q,y) )\,\partial_Q g(Q,y) + T \partial^2_{QQ}g(Q,y) 
\]
Finally, the noise $\eta_t$ has zero mean and stationary  covariance 
\begin{equation}
\label{noi}
\langle \eta_0 \, \eta_s\rangle = \langle g(Q_0,y)\,;\, g(Q_s,y) \rangle^y
\end{equation}
of order $\varepsilon^2$, reproducing the first term of \eqref{fric}.  When the sea dynamics \eqref{pertu} is undriven ($F=0$) then its stationary density $\rho_y \propto \exp -\beta U(Q,y)$ is given by the Boltzmann weight and detailed balance implies
\[
\langle Lg(Q_u,y) \,g(Q_s,y) \rangle^y_{\text eq}  = -\frac{\id}{\id u}\langle g(Q_u,y) \,g(Q_s,y) \rangle^y_{\text eq}, \quad u<s
\]
so that under these equilibrium conditions the standard second fluctuation-dissipation relation holds true:
\[
 \gamma^{\text eq}(s) = \beta\,\langle g(Q_0,y)\,;\, g(Q_s,y) \rangle^y_{\text eq} = \beta\langle \eta_0 \, \eta_s\rangle
\]
As we learn from \eqref{fric}, for nonequilibrium baths ($F\neq 0$) that relation changes into
\begin{equation}
\label{nfr}
\gamma(t) + \gamma_+(t) = \beta\, \langle \eta_t \eta_0\rangle
\end{equation}
for nonequilibrium correction
\[
\gamma_+(t) = \frac{\beta}{2}\big[\langle g(Q_0,y)\,;\, g(Q_t,y) \rangle^y + \int_{-\infty}^0\id u \langle Lg(Q_u,y) \,g(Q_t,y) \rangle^y\big],\quad t\geq 0
\]

It is important to emphasize how that new relation \eqref{nfr}  fundamentally differs from its equilibrium version.  As emphasized before in other contexts \cite{fdr,fdr1,up,fren}, we see the addition of a frenetic contribution to the friction.  The function $Lg$ is directly related to the change in dynamical activity of the sea by a change in probe position. That will be explained in Section \ref{nfdr}. The ``entropic'' contribution $\langle g(Q_0,y)\,;\, g(Q_t,y) \rangle^y$ to the friction is in terms of the autocorrelation function of the change in energy by the probe's position, and suffices under equilibrium conditions.  It is in that way that we see how nonequilibrium adds more kinetic factors to the (equilibrium) thermodynamic considerations.\\

If we take the important case of linear coupling to the sea-positions $g(Q,q) = \lambda(q-Q)$, then $Lg(Q,q) = -\lambda F(Q) + O(\lambda^2)$ is for weak coupling equal to the force on the sea particles. Measuring the friction will thus give information about that force through the correlations $\langle F(Q_u) (Q_s-y) \rangle^y$ in the friction kernel.  That resembles the situation for the extended Sutherland-Einstein relation between diffusion and mobility \cite{proc,sog,pro}. More generally, we pick up correlations between dynamical activity and force.  

\section{Second equilibrium relation}\label{2nd}
For practical purposes, the second fluctuation--dissipation theorem is mostly a physically motivated modeling assumption in diffusion processes.  More fundamentally it says something physically interesting about reduced dynamics: if that reduced dynamics for a particle in contact with equilibrium baths takes the form of a (generalized) Langevin equation, then the noise-covariance equals the memory kernel in the friction up to a factor $k_BT$ of thermal energy.  In that sense, applying the standard second fluctuation--dissipation relation between friction and noise, one effectively restricts the nonequilibrium aspect to the probe, which then alone carries the only degrees of freedom in the universe which are being driven.

The standard  second fluctuation-dissipation theorem  requires an assumption of equilibrium for the reservoirs, and/or combined with a weak coupling assumption between probe and environment; see e.g. the introduction of \cite{har} for a good understanding.    We  very briefly repeat here three ways in situations of increasing complexity of deriving that standard equilibrium relation. 

\subsection{From equipartition}
The simplest and best known derivation of the relation between noise and friction starts from a Markov diffusion process and requires that the stationary distribution must be the appropriate Gibbs distribution, in particular showing the correct Maxwellian velocity distribution.  We are in full open equilibrium with say a Langevin particle suspended in a single heat bath,
\[
\dot{x}=v,\quad \dot{v} = -V'(x)  - \gamma v + \eta
\]
(taking, for simplicity, real position $v$, mass one and velocity $v$.)
We assume that the Gaussian noise $\eta$ is white and we are asked to find its amplitude $2D = \langle \eta^2\rangle$ to ensure that the corresponding equilibrium has density $\rho(x,v) \sim \exp[-(v^2/2+V(x))/k_BT]$.  We can insert that condition in the stationary Fokker-Planck equation
\[
v\,\frac{\partial}{\partial x} \rho(x,v) + \frac{\partial}{\partial v}[\{V'(x)+\gamma v \} \rho(x,v)]
+ D\,\frac{\partial^2}{\partial v^2} \rho(x,v) = 0
\]
to find $D =\gamma k_BT$.  That means that the diffusion in velocity space is proportional to the friction.  That way of reasoning becomes even easier when the stationary velocity variance is available.  It is then sufficient to require $\langle v^2\rangle = k_BT$
for finding the amplitude $D$, whence the title of this subsection which refers to the keyword in many by now standard derivations following the original Kubo treatment \cite{kubo,tod92}.

\subsection{From local detailed balance}
The previous argument ``from equipartition'' can be extended to cases where we do not know the stationary distribution, by imposing the condition of local detailed balance.  We start now from a generalized Langevin description for the position $x_t$ and the velocity
$v_t$ of a (mass 1) particle:
\begin{eqnarray}\label{vin}
\frac{\id x_t}{\id t} & = & v_t  \\ \frac{\id v_t}{\id t} & =
& - \int_{-\infty}^t \id s \; \gamma(t-s) v_s + F_t(x_t) +
\eta_t 
\nonumber
\end{eqnarray}
We use a one-dimensional notation for simplicity even though higher dimensions are in general necessary to 
accommodate non-conservative forces $F_t$. The friction memory kernel $\gamma(s)$ is non-negative but vanishes
for negatives times $s<0$.  The $F_t$ is the possibly time-dependent forcing and the noise $\eta_t$ is assumed drawn from  a mean zero stationary 
Gaussian process. We ask for the relation between $\gamma(s)$ and the noise covariance $\langle \eta_s\eta_0\rangle$.\\

The answer is that correct modeling of motion in interaction with separate equilibrium baths requires that the noise $\eta_t$ be so related to the friction kernel $\gamma$ in \eqref{vin} that the path-wise entropy flux (per $k_B$) equals the source term of time-reversal breaking, which is the condition of local detailed balance
\begin{equation}\label{db1}
\frac{\mbox{Prob}[\omega]}{\widetilde{\mbox{Prob}}[\theta\omega]} = \exp \big(\frac 1{k_B} \mbox{ total entropy flux in } \omega\big)
\end{equation}
for every system path $\omega$ with $\theta \omega$ its time-reversal, and where $\widetilde{\mbox{Prob}}$ is the probability under reversed protocol of the dynamics.
The total entropy flux is the time-integrated entropy flux in all equilibrium reservoirs as seen from the path $\omega$ of the system. 
That basic modeling assumption was first described in \cite{lebo,ka}, see also Section 2 in \cite{der}, but the deeper reason for local detailed balance is the microscopic time-reversibility as extended to systems consecutively in contact with different equilibrium reservoirs --- see \cite{mae03,hal,har}.   We illustrate with just the above  example \eqref{vin}  how that requirement \eqref{db1} also leads to the standard second fluctuation--dissipation relation.\\

Assuming an equilibrium medium at uniform temperature $T$, the entropy flux per $k_B$ for the model \eqref{vin} is 
\begin{equation}\label{ldb}
\frac 1{k_BT}\{- \int \id s \, \dot{v}_s\,v_s +  \int \id s \, F_s(x_s)\,v_s\}
\end{equation}
The first term in the right-hand side is a temporal boundary term
accounting for the kinetic energy difference between the initial and
final state of the trajectory.  The second term refers to the time-integrated dissipated power by the forcing $F_t$.
The expression \eqref{ldb} specifies the right-hand side of \eqref{db1}.  The left-hand side of \eqref{db1} follows from a path-integration formula where time-reversal and reversed protocol are defined as
\begin{equation}
 \theta x_t  =  x_{-t},\quad
\theta v_t  =  -v_{-t}\quad
F_t \rightarrow  F_{-t} 
\end{equation}
 When the $\eta_t =(\eta_t^i)$ would be multidimensional we also assume that the noise is time-reversal invariant in the sense that $\langle \eta^i_t\eta^j_0\rangle = \langle \eta^j_t\eta^i_0\rangle$.  The rest is a computation of stochastic calculus as e.g. done in the Appendix of \cite{sog},  to conclude that local detailed balance \eqref{db1}
is verified whenever
\begin{equation}
\langle \eta_s \eta_t\rangle = k_BT\,\gamma(|t-s|)
\label{loc_det_bal}
\end{equation}
between the noise covariance and the symmetric part of the memory
kernel.  We see that the obtained relation \eqref{loc_det_bal} is as such independent of the nonequilibrium driving $F_t$  as it just expresses the thermal equilibrium of the bath.  In contrast, for the present paper, any driving will be applied directly on the intermediate bath particles (= the sea) in interaction with the probe.

\subsection{From the first fluctuation--dissipation theorem}\label{f1}
The usual derivation of the second fluctuation-dissipation relation from the first one goes by assuming a (generalized) Langevin equation for the probe and by requiring that the Kubo formula holds for linear response, \cite{kubo}.  
Yet, for probes in contact with nonequilibrium (active) baths, we do not know the physics of linear response and there is no alternative to first investigating what friction--noise relation emerges in a reduced dynamics.  For preparing that strike in Section \ref{n2} for active baths we give already here  that alternative in the detailed balance case, where a generalized Langevin equation, including the standard relation between friction and noise for equilibrium baths, is actually derived using linear response around equilibrium.\\

We illustrate the procedure for a colloidal particle in a bath of particles; see Section 1.6 of Zwanzig's book, \cite{zwanzig} for an exact calculation with independent bath particles.\\
The probe (or colloid) of mass $M$ is described by a real coordinate $q$ and moves in a potential $V$ and in (harmonic) contact with bath particles. One should think of it moving on a much slower time-scale than the bath particles which are  described by a set of coordinates $Q = \{q^j\}$ with interaction potential $\Phi(q^j-q^{j'}) = \Phi(q^{j'}-q^j)$, and moving themselves in an equilibrium fluid at temperature $T$.\\ 
The coupled equations of motion are then taken to be 
\begin{eqnarray}\label{hami}
M\frac{\id^2 q_t}{\id t^2} &=& 
-V'(q_t) + \,\sum_j \lambda_j\varepsilon_j\,[q^j_t-\varepsilon_j\,q_t], \quad q_0=y,\quad  V'(y)=0\nonumber\\
\frac{\id q^j_t}{\id t} &=& 
-\sum_{j'\neq j}\Phi'(q^j_t-q^{j'}_t) - \lambda_j\,[q^j_t-\varepsilon_j\,q_t] + \sqrt{2k_BT}\,\xi^j_t\\
 &=& -\sum_{j'\neq j}\Phi'(q^j_t-q^{j'}_t)- \lambda_j\,[q^j_t-\varepsilon_j\,y]  + \varepsilon_j  \lambda_j \,[q_t-y] + \sqrt{2k_BT}\,\xi^j_t
\nonumber
\end{eqnarray}
where the last terms contain the standard white noises $\xi^j_t$, independent over the bath particles.
We take $y$ to be the equilibrium probe position.\\
The energy function for the bath particles is
\[
U(Q,q) =  \sum_{j<j'}\Phi(q^j-q^{j'}) + \sum_j\frac{ \lambda_j}{2}\,(q^j-\varepsilon_j q)^2
\]
with coupling $ - \partial_q U(Q,q) = \sum_j \varepsilon_j  \lambda_j\,(q^j-\varepsilon_j q)$ of order $\varepsilon_j = O(\varepsilon)$.  We expand in the $\varepsilon_j$, first for the statics
\begin{equation}\label{stap}
\frac 1{Z_q}\int X^\varepsilon(Q)e^{-\beta U(Q,q)} = \langle X^\varepsilon \rangle^y + \beta(q-y)
\langle X^\varepsilon ; X^\varepsilon \rangle^y,\quad Z_q=\exp -\beta {\cal F}(q)
\end{equation}
where we use the sea-observable $X^\varepsilon(Q) = \sum_i \varepsilon_i\,\gamma_i\ q^i$, and
$\langle \cdot \rangle^y$ is in the  Gibbs distribution with expectations 
\begin{equation}\label{cano}
\langle f(Q) \rangle^y := \frac 1{Z_y}\int(\prod_j\id q^j)\,f(\{q^j\})\,e^{-U(\{q^j\},y)/(k_BT)}
\end{equation}
which is left invariant by the (unperturbed) dynamics
\begin{equation}\label{unp}
\frac{\id q^j_t}{\id t} = 
-\sum_{j'\neq j}\Phi'(q^j_t-q^{j'}_t) - \lambda_j\,[q^j_t-\varepsilon_j\,y] + \sqrt{2k_BT}\,\xi^j_t
\end{equation}
being the reference dynamics for \eqref{hami} when fixing the position $y$ of the probe.  We also start at $q_0=y$.\\
Secondly we use the Kubo formula for the dynamics.
The third term  on the right in the last line of \eqref{hami} (proportional to $\varepsilon_j$) is the time-dependent perturbation; on the time-scale of the bath the term $\varepsilon_j(q_t-y)$ is very small.  We thus apply  the (dynamical) linear response for the expectations $\langle \cdot\rangle$ in the perturbed process \eqref{hami}, started at time $t=0$ from the equilibrium ensemble \eqref{cano}. We write $X^\varepsilon(Q_t) = X^\varepsilon_t$. The Kubo formula for linear response gives to $O(\varepsilon^2)$,
\begin{equation}\label{skubo}
\langle X^\varepsilon_t\rangle= \langle X^\varepsilon \rangle^y + \beta\int_0^t\id s\,[q_s-y]
\frac{\id}{\id s}\langle X^\varepsilon_s\, X^\varepsilon_t \rangle^y
\end{equation}
in which the integral by partial integration becomes
\begin{eqnarray}\label{sup}
\int_0^t\id s\,[q_s-y]
\frac{\id}{\id s} \langle X^\varepsilon_s X^\varepsilon_t \rangle^y &=& -\int_0^t\id s\,\dot{q}_s\,
\langle X^\varepsilon_s ; X^\varepsilon_t \rangle^y \nonumber\\
+ && [q_t-y]\, \mbox{ Var}^0 X^\varepsilon
\end{eqnarray}
in terms of the connected correlation function $\langle X^\varepsilon_s ; X^\varepsilon_t \rangle^y  = \langle X^\varepsilon_s  X^\varepsilon_t \rangle^y  - \langle X^\varepsilon_s  \rangle^y\,\langle X^\varepsilon_t \rangle^y$.\\
Now, from \eqref{stap} we can substitute in \eqref{sup},
\begin{equation}\label{supper}
\langle X^\varepsilon \rangle^y
+  \beta[q_t-y]\, \mbox{ Var}^0 X^\varepsilon = \frac 1{Z_{q_t}}\int X^\varepsilon(Q)e^{-\beta U(Q,q_t)}
\end{equation}
with the appearance of the statistical force
\[
\frac 1{Z_{q}}\int X^\varepsilon(Q)e^{-\beta U(Q,q)} - \sum_j \varepsilon_j^2  \lambda_j q =k_BT\,\partial_q \log Z_q
\]
Continuing with \eqref{skubo}, we thus have
\[
\langle X^\varepsilon(t)\rangle  = \frac 1{Z_{q_t}}\int X^\varepsilon(Q)e^{-\beta U(Q,q_t)} - \frac{1}{k_BT}\int_0^t\id s\,\dot{q}_s\,\langle X^\varepsilon_s ; X^\varepsilon_t \rangle^y
\]
We now go back to the equation \eqref{hami} for $q_t$, in which we put
\[
X^\varepsilon_t = \langle X^\varepsilon_t\rangle + \eta(t)
\]
to get
\begin{equation}\label{col}
M\frac{\id^2 }{\id t^2}q_t = G(q_t) -\frac{1}{k_BT}\int_0^t\id s\,\dot{q}_s\,
 \langle X^\varepsilon_s; X^\varepsilon_t \rangle^y + \eta_t -V'(q_t)
 \end{equation}
for statistical force (derived from the equilibrium free energy $\cal F$)
\[
G(q) = -\partial_q {\cal F}(q), \quad {\cal F}(q) = -k_BT\,\partial_q \log Z_q
\]
and where the noise $\eta_t$ has mean $\langle \eta_t \rangle =0$, and covariance
 \begin{equation}\label{sec}
\langle \eta_t\eta_s\rangle  = \langle X^\varepsilon_s; X^\varepsilon_t \rangle^y
\end{equation}
to leading order as in \eqref{skubo}.  The identity \eqref{sec} in the effective (reduced) dynamics 
\eqref{col} for the probe or colloid is again the (standard) second fluctuation--dissipation theorem.  Note that $X^\varepsilon$ is a macroscopic observable in the sea degrees of freedom, which for a suitable choice of  scaling of the $\varepsilon_j,  \lambda_j$ will make the noise Gaussian and white.

\section{Nonequilibrium linear response}\label{nfdr}
Recent years have seen a growing interest in understanding the linear response around nonequilibrium.  A review with different points of view and different mathematical approaches have been presented in \cite{up}.  Here we choose for the path-space approach whose physical interpretation for linear response was especially emphasized in \cite{fdr}. There is a reference process with expecations denoted by $\langle \cdot\rangle^{\text{ref}}$.  Starting from the same initial condition at time zero a perturbation is switched on.  The general structure in the change of an observation of $O$ over time $[0,t]$ is
\begin{equation}\label{dec2}
\langle O\rangle - \langle O\rangle^{\text{ref}} = 
  \frac 1{2}\langle \mbox{Ent}^{[0,t]}(\omega)\,O\rangle^{\text{ref}}
- \langle \mbox{Esc}^{[0,t]}(\omega)\,O\rangle^{\text{ref}}
\end{equation}
where Ent$^{[0,t]}(\omega)$ is the excess in entropy flux over the time period $[0,t]$ and 
Esc$^{[0,t]}(\omega)$ is the excess in dynamical activity due to the perturbation.  When $O = A_t$ only observes at the single time $t$, then \eqref{dec2} is
  \begin{equation}\label{dec3}
\langle A_t\rangle - \langle A_t\rangle^{\text{ref}} = 
 \frac 1{2}\langle \mbox{Ent}^{[0,t]}(\omega)\,A_t\rangle^{\text{ref}}
- \langle \mbox{Esc}^{[0,t]}(\omega)\,A_t\rangle^{\text{ref}}
\end{equation}
When dealing with standard equilibrium averages $\langle \cdot \rangle^{\text{ref}} = \langle \cdot \rangle_{\text{eq}}$, we have the further identity that
\[
\langle \mbox{Esc}^{[0,t]}(\omega)\,A_t\rangle_{\text{eq}} = \sigma_A\,\langle \mbox{Esc}^{[0,t]}(\omega)\,A_0\rangle_{\text{eq}} = \frac 1{2}\sigma_A\langle \mbox{Ent}^{[0,t]}(\omega)\,A_0\rangle_{\text{eq}} = -\frac 1{2}\langle \mbox{Ent}^{[0,t]}(\omega)\,A_t\rangle_{\text{eq}} 
\]
where $\sigma_A = \pm 1$ is the parity of observable $A$ under kinematic time-reversal.   As a consequence then, under stationary time-reversibility, \eqref{dec2} reduces to the standard fluctuation--dissipation theorem and classical Kubo formula.
Away from equilibrium, the second term in the response formula \eqref{dec2}, the so called frenetic contribution, is independent and essential ; see e.g, the frenetic origin of negative differential response in \cite{fren} and the modification of the Sutherland--Einstein relation in \cite{proc,sog,pro}. We illustrate the situation for Langevin dynamics.\\

We consider a perturbation in the form of a potential $U(Q)$.
The state of the particles is $(Q,P) =
(q^1,q^2,\ldots,q^n;$ $p^1,p^2,\ldots,p^n)\in \bbR^{2n}$, collecting positions
and momenta of degrees of freedom to which is assigned a standard white noise $\xi^i_t$ with constant strength $D^i$ and a
friction coefficient $\gamma^i =D^i/T^i$:
\begin{eqnarray}
\dot{q}^i &=& p^i \nonumber\\
\dot{p}^i &=&
 F^i(q)  -\gamma^i p^i  +
  h_t \frac{\partial U}{\partial q^i}  + \sqrt{2D^i}\,\xi^i_t
\label{ud}
\end{eqnarray}
They are modeling the sea particles with which the probe will interact.
The forces $F^i$ work on these particles and are supposed to contain a nonconservative part.  For the existence of a stationary distribution one would need that the forces are sufficiently confining  so that the particles typically reside in a bounded region.  We already inserted the perturbation $U(Q)$ with small
 time-dependent amplitude $h_s$ for $ s\geq 0$.  The linear response is given by \eqref{dec3} which now becomes $\langle A_t \rangle - \langle A  \rangle^{\text{ref}} = \int_0^t \id s R(t,s) h_s$ with susceptibility
 \begin{eqnarray}\label{genfor}
 R(t,s) =& \sum_{i}\frac 1{2T^i}&\!\!\!\!\!\!\!\!\!
\langle\frac{\partial U}{\partial q^i}(Q_s) \,p^i_s\,A_t \rangle^{\text{ref}} \nonumber\\ 
&-\sum_i\frac 1{2D^i} \Bigg\{&
\langle \frac{\partial U}{\partial q^i}(q_s)\,F^i(Q_s)\,A_t \rangle^{\text{ref}}
 - \frac{\id}{\id s} \langle\frac{\partial U}{\partial q^i}(Q_s)\, p^i_s\,A_t \rangle^{\text{ref}}\nonumber\\
 &&
 + \sum_j \langle \frac{\partial^2 U}{\partial q^j\partial q^i}(Q_s)\,p^j_s\,p^i_s\,A_t \rangle^{\text{ref}}
\Bigg\}
 \end{eqnarray}
  We refer to formula (17) 
in \cite{fdr1} for the detailed derivation.  We will actually only need the formula \eqref{genfor} in the overdamped case  for $\gamma^i=1, D^i=T^i=T$ so that we truly deal with sea-evolution
\[
\dot{q}^i =
 F^i(Q) +
  h_t \frac{\partial U}{\partial q^i}(Q)  + \sqrt{2T}\,\xi^i_t
  \]
  in which case the linear response becomes
  \begin{equation}\label{linrt}
\langle A_t\rangle - \langle A \rangle^{\text{ref}} =  \frac{\beta}{2} \int_0^t \id s\, h_s\, \big[ \frac{\id}{\id s} \langle U(Q_s)\,A_t \rangle^{\text{ref}} - \langle LU(Q_s)\,A_t \rangle^{\text{ref}}\big]
  \end{equation}
  for backward generator $LU = \sum_i [F^i \partial_{q^i} U + T \partial^2_{q^iq^i} U]$.
To see the change in the stationary density $\rho$  we should take in \eqref{linrt} the amplitude $h_s\equiv h$ constant and let $t\uparrow \infty$, to obtain
\begin{equation}\label{stku}
\langle A \rangle - \langle A \rangle^{\text{ref}} = 
\frac{\beta\,h}{2}\big[ \langle U;A \rangle^{\text{ref}} - \int_{-\infty}^0\id u \langle LU(Q_u)\,A_0 \rangle^{\text{ref}}\big]
  \end{equation}
  The perturbation formul{\ae} \eqref{linrt} and \eqref{stku} will replace the equilibrium expansions
  \eqref{skubo} and \eqref{stap} in the nonequilibrium situation of next section.  For applications and for the discussion of the result in Section \ref{disc}{ it is important to repeat the interpretations that connect with the symbols in \eqref{dec2}--\eqref{dec3}.
  The first sum in \eqref{genfor} and the first term in \eqref{linrt} correspond to the dissipative part from the entropy fluxes in the thermal reservoirs; that is the ``Ent'' part.   The remaining sums and term give the frenetic contribution. They correspond to the excess in dynamical activity caused by the perturbation. As these are related to the escape rate we use the symbol ``Esc.''  A more precise connection is for example contained in Appendix A of \cite{mean}.

\section{Second nonequilibrium relation}\label{n2}
We come to the main point of the paper, to use the nonequilibrium relations of the previous section, in particular \eqref{dec3} in the forms \eqref{stku}--\eqref{linrt}, for applying it to models like in Section \ref{f1} but with a nonequilibrium sea.\\

To be clear and simple on the logic of the argument we continue first in the ``one-dimensional'' notation of Section \ref{disc}.
We consider the coupled dynamics of sea $Q_t$ with probe $q_t$,
\begin{equation}\label{2pertudb}
\dot Q_t = F(Q_t) -\partial_Q U(Q_t,q_t)  + \sqrt{2T}\,\xi_t, \quad M\ddot q_t = -V'(q_t) + g(Q_t,q_t)
\end{equation}
The white noise $\xi_t$ stands for the thermal bath at temperature $T$ ($k_B=1$ now).
The evolution equation for the probe contains the coupling $g(Q,q) = -\partial_q U(Q,q)$ with the sea and needs to be integrated out.  The rest of that equation {$M\ddot q_t$ and $V'(q_t)$ will pass unchanged to the effective probe dynamics and can be chosen differently, e.g. in an overdamped limit.  These terms are however important for the type of limit that is considered. 
We work under the approximations  that the mass $M$ of the probe is sufficiently big and the coupling  is sufficiently small to expand in both coupling and time-scales consistently.   As in the equilibrium case of Section \ref{f1} we assume that there is a unique $y$ with $V'(y)=0$ which is the most likely position of the probe and we put it also there initially, $q_0=y$. \\
For each fixed probe position $q$ we assume there is a unique and smooth stationary density $\rho_q(Q)$ for the sea, but we do not know its form except perturbatively through \eqref{stku}:
\begin{equation}\label{2pf}
\int \id Q\,\rho_q(Q)\,A(Q) = \langle A \rangle^y + \frac{\beta}{2}\,(q-y)\,\big[\langle g(Q,y);A(Q) \rangle^y
-\int_{-\infty}^0\id s\, \langle Lg(Q_s,y)\,A_0 \rangle^y\big]
\end{equation}
where $y$ is the preferred position of the probe, and we write $A_s = A(Q_s)$.
Our reference $\langle \cdot  \rangle^y$ is the stationary dynamics with selection of the value $y$ for the  position of the probe. The density $\rho_{y}$ is invariant under the reference dynamics
\[
  \dot Q_t = F(Q_t) -\partial_Q U(Q_t,y)  + \sqrt{2T}\,\xi_t
\]
The time-dependent perturbation formula \eqref{linrt} for the sea-dynamics applies  as
\begin{equation}\label{2stf}
\langle A_t\rangle = \langle A  \rangle^y + \frac{\beta}{2}\,\int_0^t\id s (q_s-y)\,\big[\frac{\id}{\id s}\langle g(Q_s,y) A_t \rangle^y - \langle Lg(Q_s,y) A_t \rangle^y\big]
\end{equation}
where the expectation in the left-hand side is with respect to the perturbed dynamics
\begin{equation}\label{pert}
\dot Q_t = F(Q_t)-\partial_Q U(Q_t,y)  - (q_t-y)\,\partial_q\partial_Q U(Q_t,y) + \sqrt{2T}\,\xi_t
\end{equation}
being the linear approximation to \eqref{2pertudb}. 
For the last term in \eqref{2stf} we do partial integration:
\begin{eqnarray}
\label{2pi}
\int_0^t\id s (q_s-y)\,\big[\frac{\id}{\id s}\langle g(Q_s,y) A_t \rangle^y - \langle Lg(Q_s,y) A_t \rangle^y\big] &=&\nonumber\\
\int_0^t\id s (q_s-y)\,\frac{\id}{\id s}\big[\langle g(Q_s,y) A_t \rangle^y - \int_{-\infty}^s\id u \langle Lg(Q_u,y) A_t \rangle^y\big] &=& \nonumber\\
-\int_0^t\id s\, \dot{q}_s\,\big[\langle g(Q_s,y)\,;\, A_t \rangle^y - \int_{-\infty}^s\id u \langle Lg(Q_u,y) A_t \rangle^y\big] &+& \nonumber\\
  (q_t -y)\big[\langle g(Q,y)\,;\,A(Q) \rangle^y-\int_{-\infty}^t \id u \langle Lg(Q_u,y)\, A_t \rangle^y\big]
  \end{eqnarray}
We can use the static response formula \eqref{2pf} to substitute the very last line via
  \begin{eqnarray}
  \frac{\beta}{2}(q_t -y)\big[\langle g(Q,y)\,;\,A(Q) \rangle^y-\int_{-\infty}^t \id u \langle Lg(Q_u,y)\, A_t\rangle^y\big] =\nonumber\\- \langle A\rangle^y +
\int\id Q\,\rho_{q_t}(Q)\, A(Q)  \label{v6}
\end{eqnarray}
 We now apply the response formula \eqref{2stf} with \eqref{2pi}--\eqref{v6} for $A(Q) \equiv g(Q,q_t) = -\partial_y U(Q,q_t)$, in order to get the effective dynamics for the probe $q_t$.\\
For the expectation of the right-hand side in \eqref{2pertudb} we get
\begin{eqnarray}\label{2fina}
\langle g(Q_t,q_t)\rangle =  \int g(Q,q_t)\rho_{q_t}(Q)\,\id Q &&\nonumber\\
- \frac{\beta}{2}
\int_0^t\id s \, \dot{q}_s\,\,\big[\langle g(Q_s,y) g(Q_t,q_t) \rangle^y &-& \int_{-\infty}^s\id u \langle Lg(Q_u,y) g(Q_t,q_t) \rangle^y\big]\nonumber\\
=  \int g(Q,q_t)\rho_{q_t}(Q)\,\id Q&&\nonumber\\
- \frac{\beta}{2}
\int_0^t\id s \, \dot{q}_s\,\,\big[\langle g(Q_s,q) g(Q_t,y) \rangle^y &-& \int_{-\infty}^s\id u \langle Lg(Q_u,y) g(Q_t,y) \rangle^y\big]
\end{eqnarray}
to significant order.
We finally define the noise
\begin{equation}\label{avep}
\eta_t = g(Q_t,q_t) - \langle g(Q_t,q_t)\rangle 
 \end{equation}
 where the average $\langle \cdot\rangle$ is  with respect to the $Q_t$-dynamics \eqref{2pertudb} for a given trajectory $q_s, 0\leq s\leq t$, started from $Q_0$ that is drawn from the stationary density $\rho_{y}$. The noise has zero mean and  its covariance is
 \begin{eqnarray}\label{cco}
\langle \eta_t  \eta_s\rangle &=& \langle g(Q_t,q_t)\,;\,g(Q_s,q_s)\rangle \nonumber\\
&=& \langle g(Q_t,y)\,;\,g(Q_s,y) \rangle^y
 \end{eqnarray}
 to quadratic order in $\varepsilon$. The effective Langevin dynamics \eqref{who} is obtained from inserting into the probe dynamics of \eqref{2pertudb} the equalities \eqref{2fina}--\eqref{cco}.\\

 To illustrate the previous formul{\ae} and the result \eqref{who} we reconsider the model in the beginning of Section \ref{disc} which resembles a popular version of the Kuramoto model. We consider plane rotators, both for the probe and for the sea degrees of freedom. The dynamics is \eqref{hamin} and $y=0$.\\
 The stationary distribution $\rho_y(Q)$ of the sea particles is actually known here but we will not use it. The interaction energy $U$ and force $g$ on the probe are
 \[
 U(Q,q) = -\lambda\sum_j \cos(q_j - q),\quad g(Q,q) = -\partial_q U(Q,q) = -\lambda\sum_j \sin(q_j - q)   
 \]
 which already determines the noise covariance.
 The backward generator $L$ for the stochastic sea dynamics while fixing the probe at $q_t=0$ acts on the force $g$ as
 \[
 Lg(Q,0) = -\lambda \sum_j \big[F^j \cos q^j  +  (\lambda + a)\sin q^j \cos q^j  -  T \sin q^j\big]
 \]
The friction kernel \eqref{kura} can now be obtained from the general expression \eqref{fric}.  The statistical force $G(q)=0$ because of the symmetry $\langle \sin \theta \rangle^0_F + \langle \sin \theta \rangle^0_{-F} =0$ for the stationary expectations for the dynamics \eqref{fav1}.

\section{Additional remarks}\label{ar}

In the previous section we have considered the case of a probe fluctuating around a fixed ``preferred'' position $y$.   In many examples of nonequilibrium baths there would however not be a preferred position but a preferred trajectory $y_s$, e.g. according to the macroscopic velocity profile of the bath.  The analysis complicates in that case, because we then want to expand around a time-dependent reference dynamics going well beyond the analysis of Section \ref{nfdr}.  In the adiabatic limit the stationary density $\rho_{y_s}$ for the sea would then be substituted in e.g. equation \eqref{2pf}. We do not give the full analysis but the result is to replace $\dot q_s$ in \eqref{who} with $\dot q_s - \dot y_s$.
All that is essential for dealing with the important problem of the dynamics of colloids placed in a 
flow.  The hydrodynamic approach there is to calculate the resistance matrix and to
use so called Faxen relations to find the particle velocity, \cite{fax}. Mesoscopic modeling
takes these hydrodynamic equations and adds noise, see e.g. \cite{radu,holz}. In fact this top-down
hydrodynamic approach is even very able to describe induced hydrodynamic interactions
between colloids suspended in a 
fluid under 
flow, with resulting structure formation; see
e.g. \cite{drien}.\\

We have not given more information about the noise than through its mean and covariance.  That would do for Gaussian noise but not otherwise.  The situation is here however again not different from equilibrium.  Extra conditions, in particular the linear coupling with an extended bath, should take care of the statistical features of the noise, from colored to white.  Formally however, the calculations above hold true even for active media consisting of few particles.  But then, mean and covariance of the noise are less relevant.  The formal structure of the reduced probe dynamics remains intact.\\

A further question regarding effective dynamics is the study of fluctuation-induced forces {\it between} different probes that are immersed in the active bath.  That has been studied in a great variety of contexts such as for nonequilibrium Casimir forces \cite{mat} or for depletion forces \cite{depl} etc., but the simpler case of the (nonequilibrium) hydrodynamic interaction between several Brownian particles also still awaits further clarifications \cite{sas,carl,drien}.  It appears for example that the debate about the validity of Newton's third (action--reaction) law
has not been completely settled yet, especially in the absence of a unique definition of ``statistical'' or ``effective'' force \cite{hay,depl}.  The present results show the emergence of a statistical force in the effective probe dynamics but we leave the investigation of interactions between probes to further research.\\

The result of the paper obviously has implications for the mobility of the probe, or more generally for linear response theory applied to probe motion in contact with a nonequilibrium environment.  For example, the condition of local detailed balance \eqref{db1}--\eqref{ldb} will be violated in the sense that the entropy flux must now be calculated as if the environment was at effective temperature $(\gamma + \gamma_+)T/\gamma$ with $T$ the temperature of the equilibrium heat  bath, and with $\gamma, \gamma_+$ from \eqref{fric}--\eqref{nfr}.\\

 The possibility of connecting the first with the second fluctuation--dissipation relation  has been responsible for some confusion in terminology and for different classifications, e.g. for such pioneering examples as the Sutherland-Einstein relation between mobility and diffusion and the Johnson-Nyquist relation between resistance and voltage fluctuations; what is generically called the Einstein relation can refer both to the first as to the second fluctuation--dissipation formula.  We refer to the review of Stratonovich \cite{strat} for more classification, and for a different terminology, exchanging what we call here the second with the first fluctuation--dissipation theorem.\\

 Clearly, the problem of the present paper is an old one, and various solutions have been suggested.  A good review of less recent work e.g. \cite{fox,tomita,graham} is in \cite{ey}.  In \cite{ey} is also mentioned how much of the work on these relations have a rather formal character, say on the level of manipulations with the generator and its decompositions in symmetric and antisymmetric parts. There is a also a huge literature on methods of elimination of fast variables, such as in the pioneering works of van Hove and Prigogine, of  Mori, of Zwanzig or of Van Kampen and Oppenheim, \cite{vK,KO,mac,zwanzig,zwa61,mori} for equilibrium reservoirs and mostly starting from a Hamiltonian formulation.   The same problem has been considered for nonequilibrium baths as well but traditionally an assumption of local equilibrium was made. The environment can for example  be described in terms of gradients in temperature or velocity profile; see e.g. \cite{ls,zub,per,rub}. A more systematic treatment using general local equilibrium distributions for the environment was pursued in \cite{shea,sheaopp}.  It is indeed possible to repeat much of the general projection operator techniques for obtaining a reduced description of motion. 

\section{Conclusions}
The theory of Brownian motion and of stochastic dynamics in general starts around the
derivation of the Einstein relation for motion in {\it ruhenden Fl\"ussigkeiten} (stationary liquids
in equilibrium); see the title of \cite{eins}. There is a long history of fundamental contributions on how that could be
extended towards motion in nonequilibrium and active media.  By the latter we generally mean that the 
probe (bead, colloid, polymer,...) is in direct contact with driven degrees of freedom such as  self-propelled particles or undergoing nonconservative forces, \cite{ros}.  The present work contributes in suggesting the physically correct modeling of probe dynamics in a nonequilibrium environment; when modeled via (generalized) Langevin processes, the Einstein relation should be modified in the way described in Section \ref{disc}. 
 The friction memory kernel can be decomposed into an entropic and a frenetic contribution; the noise is only connected with the entropic term.  The frenetic contribution contains the correlation between the forcing and the position of the probe.  Its measurement thus allows the reconstruction of certain properties of the active medium, cf. also \cite{pro}.\\ 
 
 \noindent {\bf Acknowledgment}:  We are grateful to Marco Baiesi and to Urna Basu for helpful discussions.\\

\end{document}